\documentclass[pdftex,twocolumn,epjc3]{svjour3}
\RequirePackage{amsmath}
\RequirePackage{mathptmx} 

\RequirePackage[T1]{fontenc}

\usepackage[dvipsnames]{xcolor}

\smartqed 

\RequirePackage{graphicx}
\graphicspath{{fig/}}
\RequirePackage{tikz}
\RequirePackage{physics}
\RequirePackage{siunitx}
\RequirePackage{flushend}
\RequirePackage[numbers,sort&compress]{natbib}
\RequirePackage[colorlinks, citecolor=blue, urlcolor=blue, linkcolor=blue, bookmarksdepth=3]{hyperref}
\journalname{Eur. Phys. J. C}

\begin{document}

\title{Prospect of undoped inorganic crystals at 77~Kelvin for low-mass dark matter search at Spallation Neutron Source}

\author{Dmitry Chernyak$^1$ \and Daniel Pershey$^2$ \and Jing Liu\thanksref{e2}$^1$ \and Keyu Ding$^1$ \and Nathan Saunders$^1$ \and Tupendra Oli$^1$}

\thankstext{e2}{e-mail: Jing.Liu@usd.edu}

\institute{Department of Physics, University of South Dakota, 414 East Clark Street, Vermillion, SD 57069, USA \and Department of Physics, Duke University, Physics Bldg., Science Dr.,Durham, NC 27708, USA}

\date{Received: date / Accepted: date}

\maketitle

\begin{abstract}
  Investigated in this work were sensitivities of a prototype detector for the detection of low-mass dark matter particles produced at the Spallation Neutron Source at the Oak Ridge National Laboratory in two years of data taking. The presumed prototype consisted of 10~kg undoped CsI or NaI scintillation crystals directly coupled with SiPM arrays operated at 77~K. Compared to the COHERENT CsI(Na) detector, a much higher light yield was assumed for the prototype. An experiment with a cylindrical 1~kg undoped CsI crystal coupled directly to two photomultiplier tubes at about 77~K was conducted as the first step to verify the idea.  A light yield of $26.0 \pm 0.4$~photoelectrons per keV electron-equivalent was achieved. This eliminated the concern of self light absorption in large crystals raised in some of the early studies.
\end{abstract}

\section{Introduction}

The weakly-interacting massive particles (WIMP)~\cite{JUNGMAN1996195}, as one of the well-motivated dark matter (DM) candidates, have been the focus of many direct dark matter search experiments in deep underground laboratories~\cite{Schumann19}. However, as the sensitivity of these canonical WIMP search experiments approach the neutrino floor~\cite{ohare20} from a few GeV to the TeV scale~\cite{lux17}, there are increasing interests to explore a broader set of DM candidates in the region below the currently focused mass range~\cite{maryland17}.

A natural generalization of the WIMP model is to drop the assumption that DM particles interact with the Standard Model (SM) matters through known forces. If a new force is responsible for the interaction between DM and SM particles, the allowed mass range of DM particles can be extended from keV to TeV scale~\cite{Kobzarev:1966qya, Blinnikov:1982eh, foot91, hodges93, berezhiani96}. This class of DM models, named hidden-sector or dark-sector DM, still share with the WIMP model a common strong motivation, that is, the thermal history of the Universe and the coupling between DM and SM particles can generate the observed DM abundance in the current Universe~\cite{PhysRevLett.39.165, Izaguirre:2015yja}.

The most widely used benchmark model in this class predicts a kinetically mixed dark photon as the new force mediator~\cite{fayet90} that couples equally to electrons and nucleons. Under this assumption, the cross section of coherent scatterings of DM particles against nuclei can be orders of magnitude higher than that of electronic scatterings. However, since sub-GeV DM particles are less efficient than heavier ones in transferring momentum to nuclei, a very low energy threshold is desired for the detection of low energy nuclear recoils. In this less explored energy region lays possibly unexpected background sources that may dampen the sensitivity of direct detection experiments deep underground.

Sub-GeV DM particles can be generated in the collision of SM particles in accelerator-based experiments through the new force mediator, or the ``portal'' particle, just as what could happen in the early hot Universe~\cite{maryland17}. They can then interact with a detector nearby through the same portal particle and get detected. Their production rate is constrained by the observed DM abundance, while their scattering rate can reveal the very nature of them. Accelerator-based experiments can therefore probe many predictive models. The majority of these models are beyond the capability of direct detection experiments, which are solely sensitive to the scattering process.

In addition to the compelling physics motivation mentioned previously, accelerator-based DM experiments also possess a technical advantage over direct detection ones, that is, the well controlled production of DM particles from a man-made source. The precise knowledge of outgoing particles, such as their energy, angular distributions and time profile, etc., can be used to suppress random radiation backgrounds that are hard to deal with in direct detection experiments.

The COHERENT experiment~\cite{coherent18} at the Oak Ridge National Laboratory (ORNL) represents a particular type of accelerator-based DM experiments, where DM particles can be produced in the decay of SM mesons generated in a proton beam dump facility, the Spallation Neutron Source (SNS), and then coherently scatter with nuclei in a detector target. A variety of detection technologies are utilized in COHERENT, including inorganic scintillating crystals, liquid noble gases and semiconductor detectors, etc. Detailed descriptions of these technologies and their complementarity in the probing of physics beyond the SM can be found in Ref.~\cite{coherent17, coherent18, lardm19}.

The technical advantage of the accelerator-based experiment can be illustrated with the COHERENT 14~kg CsI(Na) detector.  Compared to major direct DM detection experiments based on inorganic scintillators, such as DAMA~\cite{dama18}, PICO-LON~\cite{picolon16}, DM-Ice~\cite{dmice17}, ANAIS~\cite{anais19},  COSINE~\cite{cosine19}, SABRE~\cite{sabre19} and COSINUS~\cite{angloher17}, etc., the crystal used in COHERENT was not as pure in terms of its internal radioactive contamination. The passive shielding was relatively simple, and the detector was not even placed deep underground. Yet, it was used for the first observation of coherent elastic neutrino-nucleus scatterings (CEvNS) in 2017~\cite{coherent17}.  The significantly less cost and technical difficulties result from a simple fact that the narrow SNS particle generation window imposes a few orders of magnitude reduction in random radioactive background contamination.

Data taken with the 14~kg CsI(Na) detector can already be used to exclude some low-mass dark matter parameter space, which will be mentioned in a later section. However, there is a serious limitation of the detector, that is, its relatively high energy threshold of about 1~keV electron-equivalent (keVee). Lowing the energy threshold involves reducing radioactive backgrounds, instrumental noises from light sensors and crystals near the threshold, as well as increasing the detection efficiency of the system and the light yield of the crystal. Benefit from the great background suppression power at the SNS, increasing the light yield of crystals can be the main focus of the detector improvement.

The light yields of undoped NaI and CsI crystals were observed to increase rapidly when temperature goes down, and reach the highest point around 40~K~\cite{Sciver58, Nishimura95, Sailer12}. The light yields at liquid nitrogen temperature (77~K at one atmospheric pressure) are slightly lower, but for convenience, most experiments were done at about 77~K. The observed number of photons varied with the purity of crystals and light readout methods~\cite{Bonanomi52, Hahn53a, Hahn53, Sciver56, Beghian58, Sciver60, Fontana68, West70, Fontana70, Emkey76, persyk80, Woody90, Williams90, Wear96, Amsler02, Moszynski03, Moszynski03a, Moszynski05, Moszynski09, Sibczynski10, Sibczynski12, mikhailik15, csi, ess19}. Nevertheless, all measurements gave similar or higher yields than those of doped crystals at room temperature. The highest ones~\cite{Bonanomi52, Sciver56, Moszynski03, Moszynski05, ess19} almost reached the theoretical limit deduced from the band gap energy.

In 2016, one of the authors of this work measured the light yield of a small undoped CsI crystal directly coupled to a 2 inch Hamamatsu PMT R8778MODAY(AR)~\cite{yamashita10} at 80~K and achieved a yield of $20.4\pm0.8$ photoelectrons (PE) per keVee~\cite{csi}. The cylindrical crystal used in that study had a diameter of 2 inches and a thickness of 1~cm, corresponding to a mass of only 91.4~gram. Mentioned in the literature~\cite{Sciver58}, there was a concern about strong self absorption of the intrinsic scintillation light in undoped crystals, which might prevent the usage of crystals thicker than 1/2 inch from practical uses. However, later investigations revealed that the scintillation mechanism of undoped crystals~\cite{Nishimura95, mikhailik15} should be transparent to their own scintillation light. Strong absorptions mentioned in early literature may have been due to impurities in their crystals.

A cylindrical undoped CsI crystal of more than 1~kg was used to test whether the light yield would reduce as the size of the crystal increases. The experimental setup is described first. The light yield achieved with two Hamamatsu R11065 PMTs is reported secondly. After that, operational parameters of a 10~kg prototype detector at the SNS are discussed, based on which its sensitivity to detect low-mass dark matter particles is predicted last.

Due to mechanical difficulties in operating NaI crystals in cryogenic environment, the experimental investigation was done using only undoped CsI. The discussion, however, was kept generic, involving both CsI and NaI given similar scintillation properties of the two from 4 to 300~K~\cite{Sciver58, Nishimura95, Sailer12}.

\section{Light yield of cold crystal}
\subsection{Experimental setup}
\label{s:expt}
\begin{figure}[htbp] \centering
  \includegraphics[width=\linewidth]{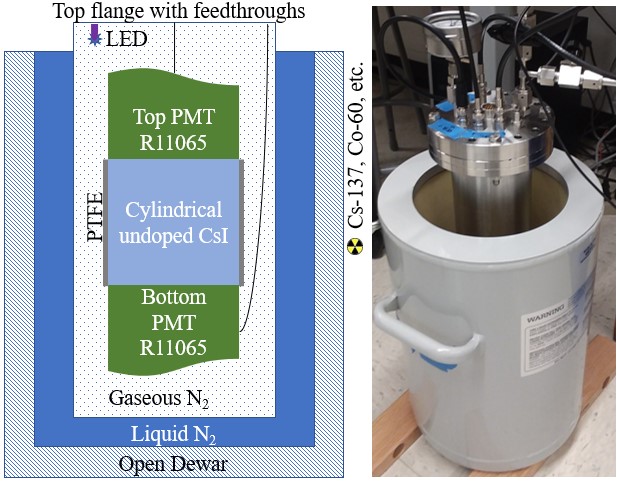}
  \caption{A sketch (left) and a picture (right) of the experimental setup.}
  \label{f:setup}
\end{figure}

The right picture in Fig.~\ref{f:setup} shows an open liquid nitrogen (LN2) dewar used to cool a 50~cm long stainless steel tube placed inside. The inner diameter of the tube was $\sim10$~cm. The tube was vacuum sealed on both ends by two 6-inch ConFlat (CF) flanges.  The bottom flange was blank and attached to the tube with a copper gasket in between. The top flange was attached to the tube with a fluorocarbon CF gasket in between for multiple operations. Vacuum welded to the top flange were five BNC, two SHV, one 19-pin electronic feedthroughs and two 1/4-inch VCR connectors.

\begin{figure}[htbp] \centering
  \includegraphics[width=\linewidth]{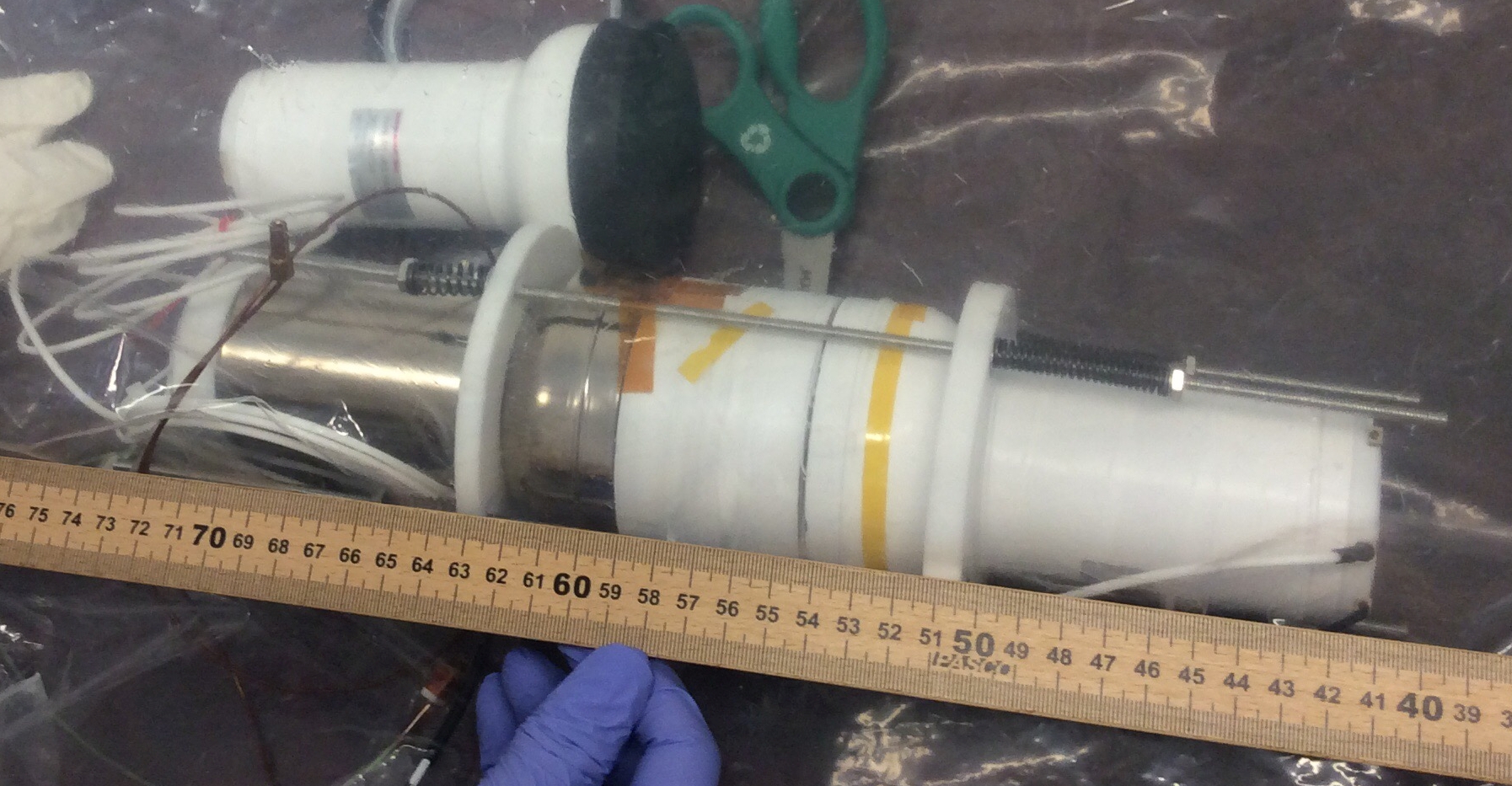}
  \caption{The detector assembly in a glove bag.}
  \label{f:det1}
\end{figure}

The left sketch in Fig.~\ref{f:setup} shows the internal structure of the experimental setup.  Three different undoped cylindrical CsI crystals were used in the measurements was purchased from the Shanghai Institute of Ceramics, Chinese Academy of Sciences. It had a diameter of 3 inches, a height of 5~cm and a mass of 1.028~kg. All surfaces were mirror polished. The side surface was wrapped with multiple layers of Teflon tapes. Two 3-inch Hamamatsu R11065-ASSY PMTs were attached to the two end surfaces without optical grease. To ensure a good optical contact, the PMTs were pushed against the crystal by springs, as shown in Fig.~\ref{f:det1}. The assembly was done in a glove bag flushed with dry nitrogen gas to minimize exposure of the crystal to atmospheric moisture. The relative humidity was kept below 5\% at 22$^{\circ}$C during the assembly process.

The assembled crystal and PMTs were lowered into the stainless steel chamber from the top. After all cables were fixed beneath it, the top flange was closed.  The chamber was then pumped with a Pfeiffer Vacuum HiCube 80 Eco to $\sim1\times {10}^\text{-4}$ mbar. Afterward, it was refilled with dry nitrogen gas to 0.17 MPa above the atmospheric pressure and placed inside the open dewar. Finally, the chamber was cooled by filling the dewar with LN2. After cooling, the chamber pressure was reduced to slightly above the atmospheric pressure.

A few Heraeus C~220 platinum resistance temperature sensors were used to monitor the cooling process. They were attached to the side surface of the crystal, the PMTs, and the top flange to obtain the temperature profile of the long chamber. A Raspberry Pi 2 computer with custom software~\cite{cravis} was used to read out the sensors. The cooling process could be done within about 30 minutes. Most measurements, however, were done after about an hour of waiting to let the system reach thermal equilibrium. The temperature of the crystal during measurements was about 3~K higher than the LN2 temperature.

The PMTs were powered by a 2-channel CAEN N1470A high voltage power supply NIM module. Their signals were fed into a 4-channel CAEN DT5751 waveform digitizer, which had a 1~GHz sampling rate, a 1~V dynamic range and a 10-bit resolution. Custom-developed software was used for data recording~\cite{daq}. The recorded binary data files were converted to CERN ROOT files for analysis~\cite{nice}.

\begin{figure}[htbp]
  \includegraphics[width=0.50\linewidth]{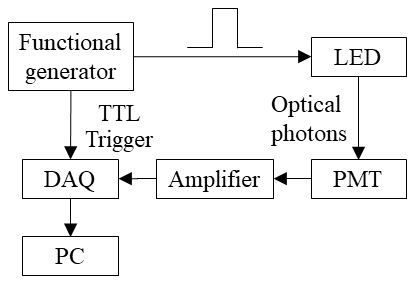}\hfill
  \includegraphics[width=0.49\linewidth]{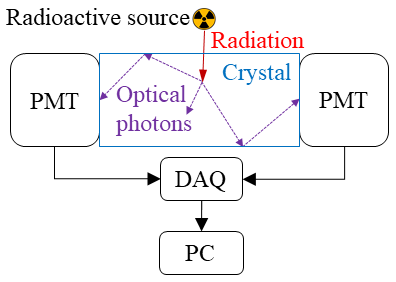}
  \caption{Trigger logics for single-photoelectron response (left) and energy calibration (right) measurements.}
  \label{f:trg}
\end{figure}

\subsection{Single-photoelectron response of PMTs}
\label{s:1pe}
The single-photoelectron response of PMTs was measured using light pulses from an ultraviolet LED from Thorlabs, LED370E. Its output spectrum peaked at 375~nm with a width of 10~nm, which was within the 200-650~nm spectral response range of the PMTs. Light pulses with a $\sim$50~ns duration and a rate of 10~kHz were generated using an RIGOL DG1022 arbitrary function generator. The intensity of light pulses was tuned by varying the output voltage of the function generator so that only one or zero photon hit one of the PMTs during the LED lit window most of the time. A TTL trigger signal was emitted from the function generator simultaneously together with each output pulse. It was used to trigger the digitizer to record the PMT response. The trigger logic is shown in the left flow chart in Fig.~\ref{f:trg}.

A typical single-photoelectron (PE) pulse from an R11065 working at its recommended operational voltage, 1500~V, is well above the pedestal noise. However, the two PMTs were operated at about 1300~V to avoid saturation of electronic signals induced by 2.6~MeV $\gamma$-rays from environmental $^{208}$Tl. The consequent small single-PE pulses hence had to be amplified by a factor of ten using a Phillips Scientific Quad Bipolar Amplifier Model 771 before being fed into the digitizer in order to separate them from the pedestal noise.

\begin{figure}[htbp]
  \includegraphics[width=\linewidth]{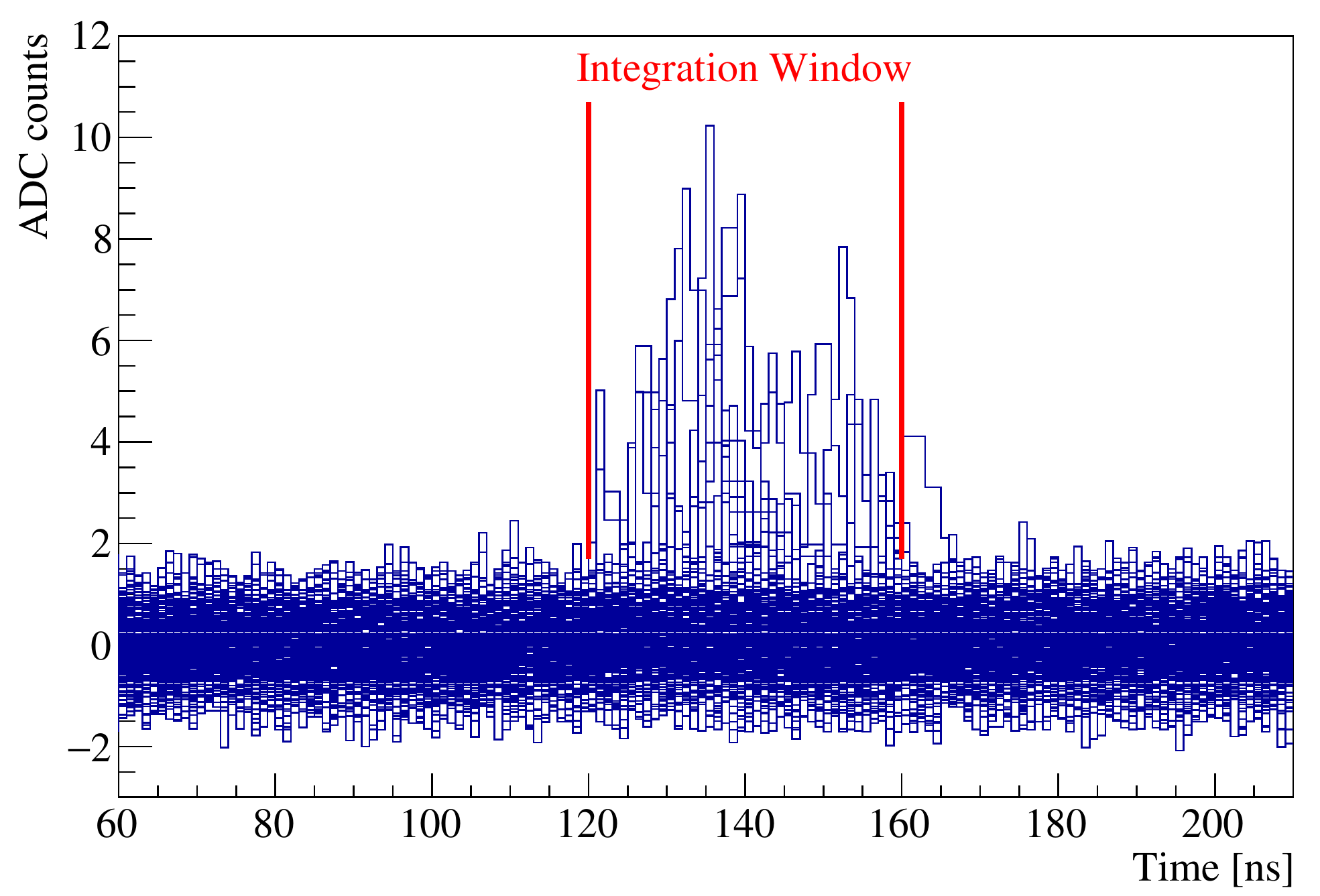}
  \caption{Two hundred consecutive waveforms from the bottom PMT overlapped with each other.}
  \label{f:spe1}
\end{figure}

\begin{figure}[htbp]
  \includegraphics[width=\linewidth]{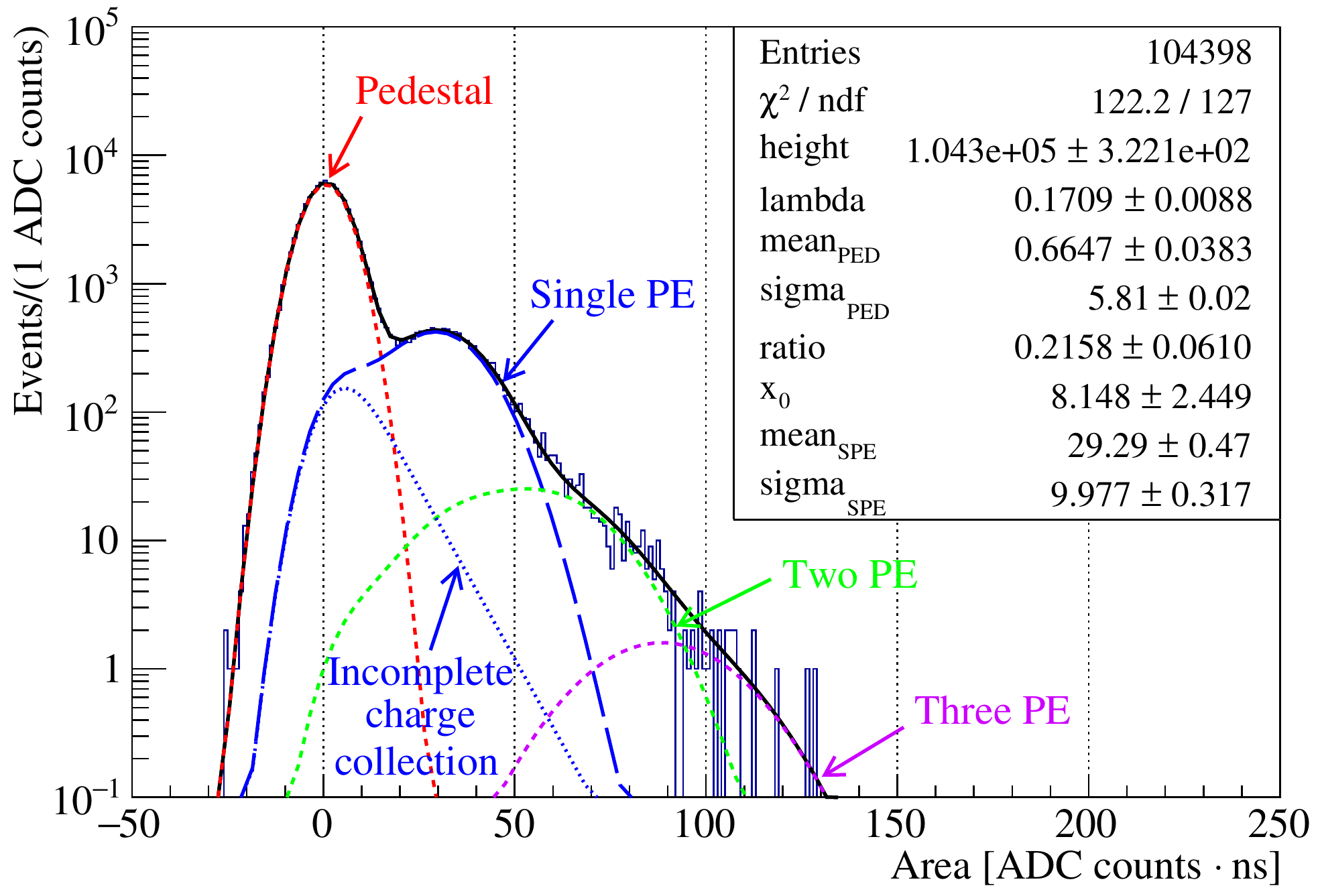}
  \caption{Single-PE response of the top PMT in logarithm scale.}
  \label{f:spe2}
\end{figure}

\begin{figure}[htbp]
  \includegraphics[width=\linewidth]{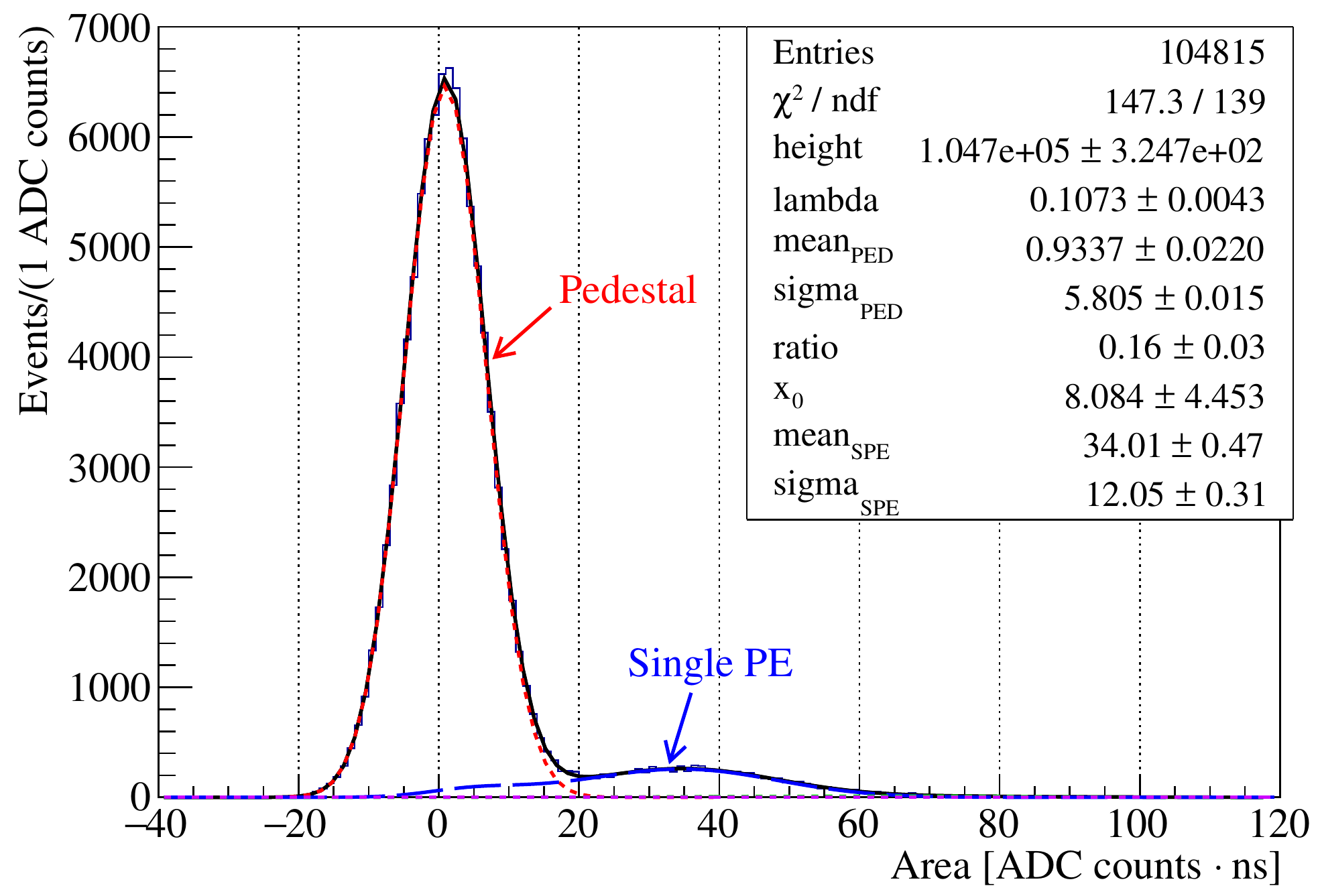}
  \caption{Single-PE response of the bottom PMT in linear scale. The two and three-PE distributions are two small to be visible.}
  \label{f:spe3}
\end{figure}

Fig.~\ref{f:spe1} shows two hundred consecutive waveforms from the bottom PMT randomly chosen from a data file taken during a single-PE response measurement. About 20 of them contain a single-PE pulse within 120 to 160~ns. An integration in this time window was performed for each waveform in the data file whether it contained a pulse or not. The resulting single-PE spectra for the top and bottom PMTs are presented in Fig.~\ref{f:spe2} and Fig.~\ref{f:spe3}, respectively.

The spectra were fitted in the same way as described in Ref.~\cite{ds1013} with a function,
\begin{equation}\label{e:Fx}
F(x)=H\sum\limits_n P(n,\lambda) f_n(x),
\end{equation}
where $H$ is a constant to match fit function to spectra counting rate, $P(n,\lambda)$ is a Poisson distribution with mean $\lambda$, which represents the average number of PE in the time window, $f_n(x)$ represents the \textit{n}-PE response, and can be expressed as
\begin{equation}\label{e:fnx}
f_n(x)=f_0(x) \ast f_1^{n\ast}(x),
\end{equation}
where $f_0(x)$ is a Gaussian function representing the pedestal noise distribution, $\ast$ denotes a mathematical convolution of two functions, and $f_1^{n\ast}(x)$ is a n-fold
convolution of the PMT single-PE response function, $f_1(x)$, with itself. The single-PE response function $f_1(x)$ was modeled as:
\begin{equation}\label{e:f1x}
f_1(x)=
  \begin{cases}
    R(\frac{1}{x_0} e^{-x/x_0})+(1 - R)G(x;\bar{x},\sigma) & x>0; \\
    0 & x\leq0,
  \end{cases}
\end{equation}
where $R$ is the ratio between an exponential decay with a decay constant $x_0$, and a Gaussian distribution $G(x;\bar{x},\sigma)$ with a mean of $\bar{x}$ and a width of $\sigma$. The former corresponds to the incomplete dynode multiplication of secondary electrons in a PMT. The latter corresponds to the full charge collection in a PMT.

The fitting result for the top PMT is shown in Fig.~\ref{f:spe2}. The fitting function has eight free parameters as shown in the top-right statistic box in Fig.~\ref{f:spe2}, where ``height'' corresponds to $H$ in Eq.~\ref{e:Fx}, ``lambda'' corresponds to $\lambda$ in Eq.~\ref{e:Fx}, ``mean'' and ``sigma'' with a subscript ``PED'' represents the mean and the sigma of the Gaussian pedestal noise distribution, those with a subscript ``SPE'' represents $\bar{x}$ and $\sigma$ in Eq.~\ref{e:f1x}, respectively, and ``ratio'' corresponds to $R$ in Eq.~\ref{e:f1x}.  Due to technical difficulties in realizing multiple function convolutions in the fitting ROOT script, the three-PE distribution, $f_1^{3\ast}(x)$, was approximated by a Gaussian function with its mean and variance three times that of the single-PE response.

Table~\ref{t:1PE} lists means of single-PE distributions for both PMTs measured before and after the energy calibration mentioned in the next section to check the stability of the PMT gains. The average mean for the top and bottom PMT is 28.58 $\pm$ 0.51 and 33.08 $\pm$ 0.47 ADC~counts$\cdot$ns, respectively.

\begin{table}[htbp]
  \centering
  \caption{\label{t:1PE} Summary of the single-PE response for the top and bottom PMTs. Top and bottom rows for each PMT correspond to measurements before and after the energy calibration, respectively.}
  \begin{tabular}{c c c c}
    \hline
    PMT &Temperature &Temperature &Mean of single-PE \\
    &of PMT [$^\circ$C]&of crystal [$^\circ$C] &[ADC counts$\cdot$ns]\\
    \hline
     Top &
     \begin{tabular}{c} -134.3 $\pm$ 1.3 \\ -134.3 $\pm$ 1.3 \\
     \end{tabular} &
     \begin{tabular}{c} -193.3 $\pm$ 1.3  \\ -193.5 $\pm$ 1.3 \\
     \end{tabular} &
     \begin{tabular}{c} 28.53  $\pm$ 0.51 \\ 28.63  $\pm$ 0.45 \\
     \end{tabular} \\
    \hline
     Bottom &
     \begin{tabular}{c} -195.7 $\pm$ 1.3 \\ -195.5 $\pm$ 1.3 \\
     \end{tabular} &
     \begin{tabular}{c} -193.3 $\pm$ 1.3  \\ -193.5 $\pm$ 1.3 \\
     \end{tabular} &
     \begin{tabular}{c} 32.98  $\pm$ 0.47 \\ 33.18  $\pm$ 0.43 \\
     \end{tabular} \\
    \hline
  \end{tabular}
\end{table}

\subsection{Energy calibration}

The energy calibration was performed using $\gamma$-rays from a $^{137}$Cs and a $^{60}$Co radioactive source, as well as $^{40}$K within the crystal and $^{208}$Tl from the environment. The sources were sequentially attached to the outer wall of the dewar as shown in Fig.~\ref{f:setup}. Background data taking was done before those with a source attached. The digitizer was triggered when both PMTs recorded a pulse above a certain threshold within a time window of 16~ns. The trigger logic is shown in the right flow chart in Fig.~\ref{f:trg}. The trigger rate for the background, $^{137}$Cs and $^{60}$Co data taking was 100 Hz, 410 Hz and 520 Hz, respectively, if the threshold was set to 10 ADC counts above the pedestal level.

\begin{figure}[htbp]
  \includegraphics[width=\linewidth]{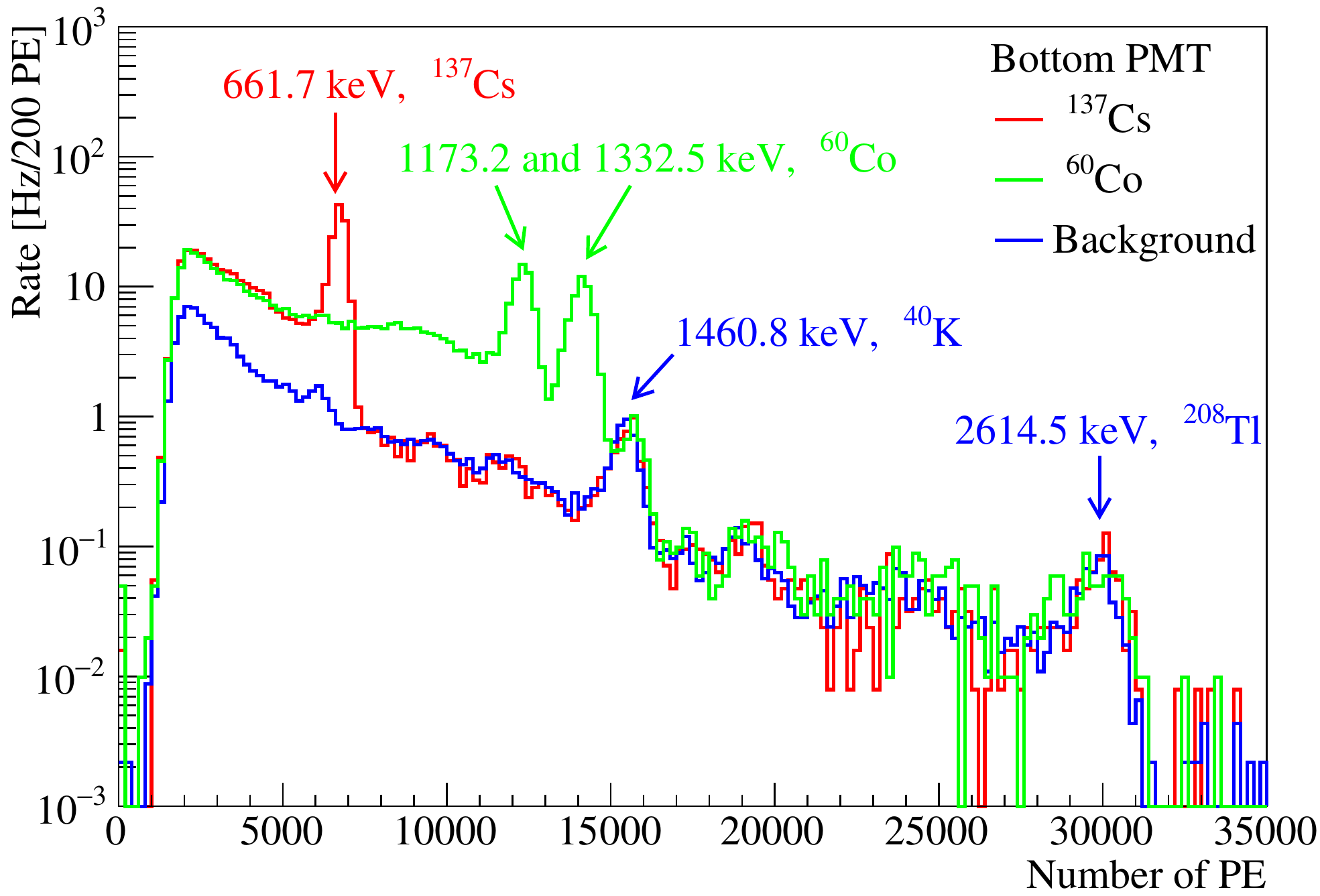}
  \caption{Energy spectra of the bottom PMT at 80~K.}
  \label{f:en1}
\end{figure}

Each recorded waveform was 8008~ns long. The rising edge of the pulse that triggered the digitizer was set to start at around 1602~ns so that there were enough samples before the pulse to extract the pedestal level of the waveform. After the pedestal level was adjusted to zero the pulse was integrated until its tail fell back to zero. The integration had a unit of ADC counts$\cdot$ns. It was converted to numbers of PE using the formula:
\begin{equation}
  \text{(number of PE)} = \text{(ADC counts} \cdot \text{ns)}/\bar{x},
  \label{e:m1pe}
\end{equation}
where $\bar{x}$ is the mean of the single-PE Gaussian distribution mentioned in Eq.~\ref{e:f1x}. Its unit was also ADC count$\cdot$ns. Its value was obtained from the fittings shown in Fig.~\ref{f:spe2} and \ref{f:spe3}.  The resulting spectra normalized by their event rates recorded by the bottom PMT are shown in Fig.~\ref{f:en1}. The spectra from the top PMT are very similar.

\begin{figure}[htbp]
  \includegraphics[width=\linewidth]{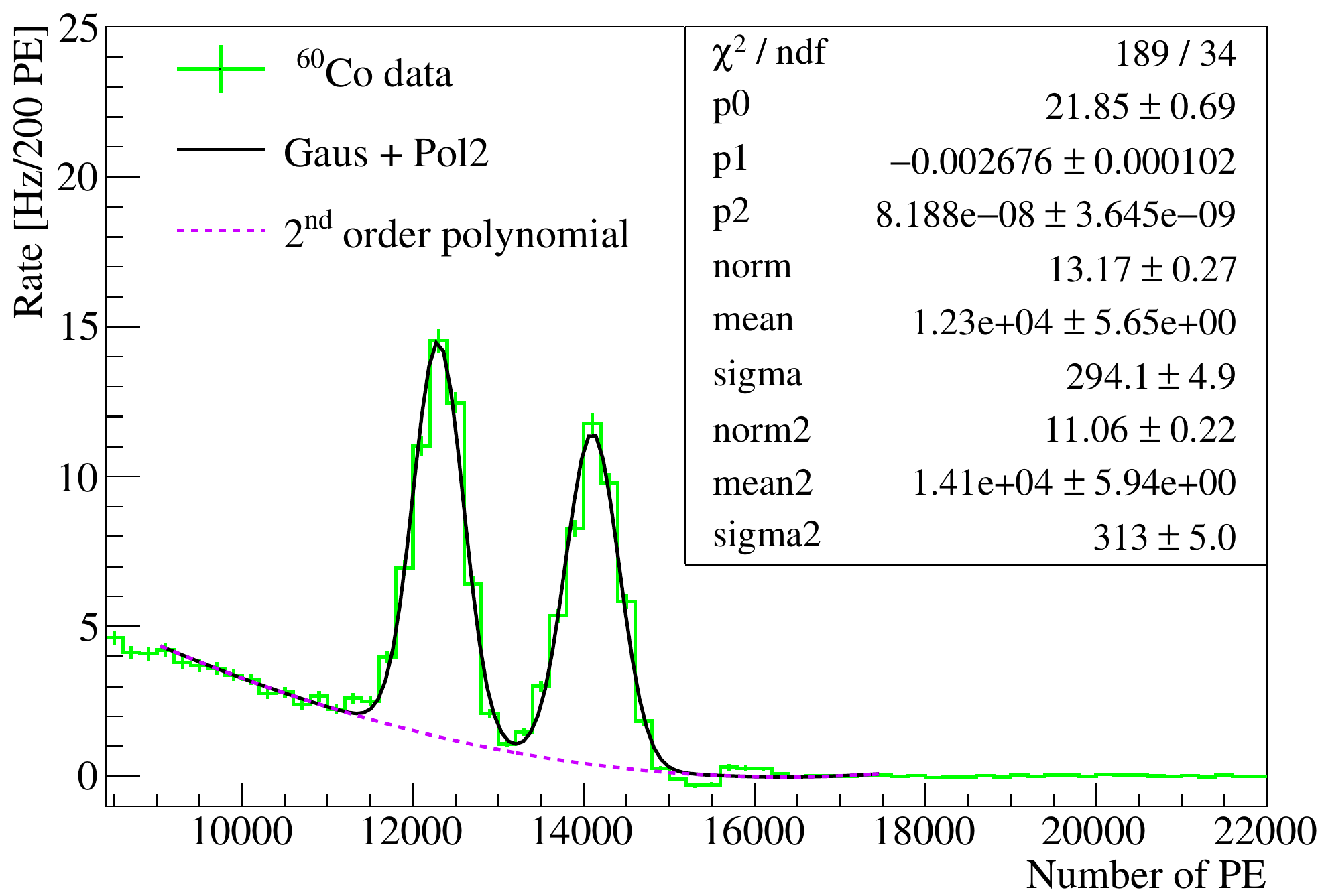}
  \caption{Energy spectrum recorded by the bottom PMT at 80~K with a $^{60}$Co source. The clearly separated peaks were fitted simultaneously with two Gaussian distributions on top of a 2$^{nd}$ order polynomial.}
  \label{f:co60}
\end{figure}

The $\gamma$-ray peaks were fitted using one or two Gaussian distributions on top of a  2$^{nd}$ order polynomial. A simultaneous fit for the 1.17 MeV and 1.33 MeV peaks from $^{60}$Co is shown in Fig.~\ref{f:co60} as an example. The peaks are clearly separated indicating an energy resolution much better than that of a regular NaI(Tl) detector running at room temperature. The means and sigmas of the fitted Gaussian functions are listed in Table~\ref{t:rPE} together with those from other $\gamma$-ray peaks.

\begin{table}[htbp]
  \caption{\label{t:rPE} Summary of $\gamma$-ray peaks in the calibration spectra.}
  \begin{minipage}{\linewidth}\centering
  \begin{tabular}{cccccc}
    \hline
   PMT & Isotope & Energy & Mean & Sigma & FWHM \\
   &  & (keVee) & (PE) & (PE) & (\%) \\
    \hline
     Top &
     \begin{tabular}{r} $^{137}$Cs \\ $^{60}$Co \\ $^{60}$Co \\ $^{40}$K \\ $^{208}$Tl \\
     \end{tabular} &
     \begin{tabular}{r} 661.7 \\ 1173.2 \\ 1332.5 \\ 1460.8 \\ 2614.5 \\
     \end{tabular} &
     \begin{tabular}{r} 9314.0 \\ 18043.1 \\ 20913.6 \\ 23137.6 \\ 42723.7 \\ 
     \end{tabular} &
     \begin{tabular}{r} 338.7 \\ 552.5 \\ 579.3 \\ 645.2 \\ 611.4 \\
     \end{tabular} &
     \begin{tabular}{r} 8.6 \\ 7.2 \\ 6.5 \\ 6.6 \\ 3.4 \\
     \end{tabular} \\
     \hline
     Bottom &
     \begin{tabular}{r} $^{137}$Cs \\ $^{60}$Co \\ $^{60}$Co \\ $^{40}$K \\ $^{208}$Tl \\
     \end{tabular} &
     \begin{tabular}{r} 661.7 \\ 1173.2 \\ 1332.5 \\ 1460.8 \\ 2614.5 \\
     \end{tabular} &
     \begin{tabular}{r} 6716.1 \\ 12297.2 \\ 14103.6 \\ 15416.5 \\ 29758.1 \\ 
     \end{tabular} &
     \begin{tabular}{r} 204.2 \\ 294.1 \\ 313.0 \\ 331.6 \\ 483.0 \\
     \end{tabular} &
     \begin{tabular}{r} 7.2 \\ 5.6 \\ 5.2 \\ 5.1 \\ 3.8 \\
     \end{tabular} \\
     \hline
  \end{tabular}
  \end{minipage}
\end{table}

\subsection{Light yield}
\begin{figure}[htbp]
  \includegraphics[width=\linewidth]{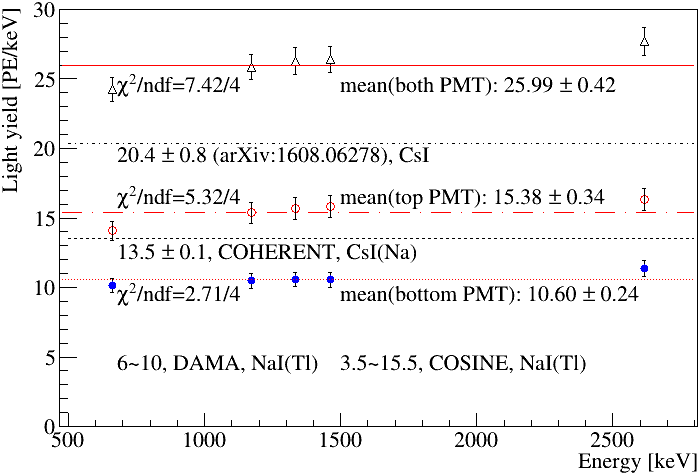}
  \caption{The obtained light yields for the top (empty circles), the bottom (filled squares) and both (empty triangles) PMTs, compared to those achieved by other experiments~\cite{coherent17, dama18, cosine19} and an earlier measurement with a smaller crystal~\cite{csi}.}
  \label{f:ly1}
\end{figure}

The light yield was calculated for each PMT using the data in Table~\ref{t:rPE} and the following equation:
\begin{equation}
  \text{light yield [PE/keVee]} = \text{Mean [PE]}/\text{Energy [keVee]}.
  \label{e:ly}
\end{equation}
The obtained light yield at each energy point is shown in Fig.~\ref{f:ly1}. The light yield of the whole system was calculated as a sum of those of the top and bottom PMTs.  The uncertainties of light yields are mainly due to the uncertainties of mean values of the single-PE responses used to convert the \textit{x}-axes of the energy spectra from ADC counts$\cdot$ns to the number of PE.  The data points in each category were fitted by a straight line to get an average light yield, which was 15.38 $\pm$ 0.34 PE/keVee for the top PMT, 10.60 $\pm$ 0.24 PE/keVee for the bottom one, and 25.99 $\pm$ 0.42 PE/keVee for the system.

To understand the origin of the significant light yield difference between the two PMTs, additional measurements were performed. First, the PMT-crystal assembly were pulled from the chamber and reinserted upside down without any other change. The PMTs kept their yields unchanged. Second, the PMT with the lower yield was replace by another R11065. No significant change could be observed. Last, the crystal was flipped while the PMTs were kept in their original locations. Again, no significant change could be observed. Therefore, the difference in the light yields between the two PMTs was most probably due to the difference in the quantum efficiencies of individual PMTs instead of different optical interfaces or temperatures.

There seems to be a systematic decrease of the light yield as the energy decreases as shown in Fig.~\ref{f:ly1}. This may indicate some non-linearity in the energy response of the undoped CsI crystal at 77 K. If the light yield at lower energies is significantly lower. The technique under investigation may not be suitable for dark matter search. Limited by the large uncertainty of each data point, no quantitative conclusion can be drawn from this measurement. Fortunately, there exist some investigations of the non-linearity of both undoped CsI~\cite{Moszynski05} and NaI~\cite{Moszynski09} at 77~K from 5.9~keV to 1.3~MeV. The results vary with the crystals used in those studies. Some had less, others had more light yields at lower energies than that at 1.3~MeV. The difference ranges from 0 to 30\%. In order to verify the light yield of the crystal used in this study at lower energies, additional studies with an $^{241}$Am source were performed. An even higher yield was achieved in the range of [13, 60]~keV. The results will be reported in a separate publication.

\section{Prototype detector at the SNS}
In this section, some key operational parameters of the presumed prototype detector are discussed, based on which, its sensitivity to probe sub-GeV DM particles is investigated.

\subsection{Size}
The light yield achieved with the $\sim$1~kg undoped CsI is even higher than that achieved with the 91.4~gram crystal, which proves that the undoped CsI is transparent to its own scintillation light up to at least a few tens of centimeters. A 10~kg prototype can hence be made from a single crystal or a few slightly small ones, each works as a independent optical modules. The scintillation mechanism of undoped crystals is summarized here to back up this conclusion.

\begin{figure}[htbp]
  \includegraphics[width=\linewidth]{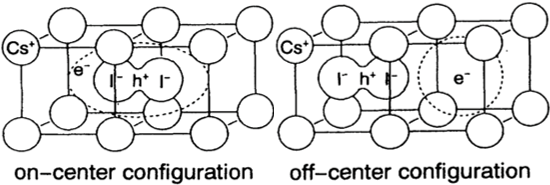}
  \caption{Two types of self-trapped excitons responsible for the scintillation emission in undoped CsI, taken from \cite{Nishimura95}.}
  \label{f:ste}
\end{figure}

A scintillation photon must have less energy than the width of the band gap of the host crystal. Otherwise, it can excite an electron from the valence band to the conduction band and be absorbed by the host crystal. This demands the existence of energy levels in between the band gap. Recombinations of electrons and holes in these levels create
photons not energetic enough to re-excite electrons up to the conduction band, and hence cannot be re-absorbed.  In Tl-doped crystals, there exist these energy levels around the doped ions, which are called scintillation centers. Scintillation centers in undoped crystals are understood to be self-trapped excitons instead of those trapped by doped impurities~\cite{pelant12}. Two types of excitons were observed in an undoped CsI~\cite{Nishimura95} as demonstrated in Fig.~\ref{f:ste}. In both cases, a hole is trapped by two negatively charged iodine ions, it can catch an excited electron and form a so-called exciton that resembles a hydrogen atom. These excitons have less energy than the width of the band gap, photons emitted by the de-excitation of which are not energetic enough to be re-absorbed by the host crystal.

\subsection{Operation temperature}
\begin{figure}[htbp]
  \includegraphics[width=\linewidth]{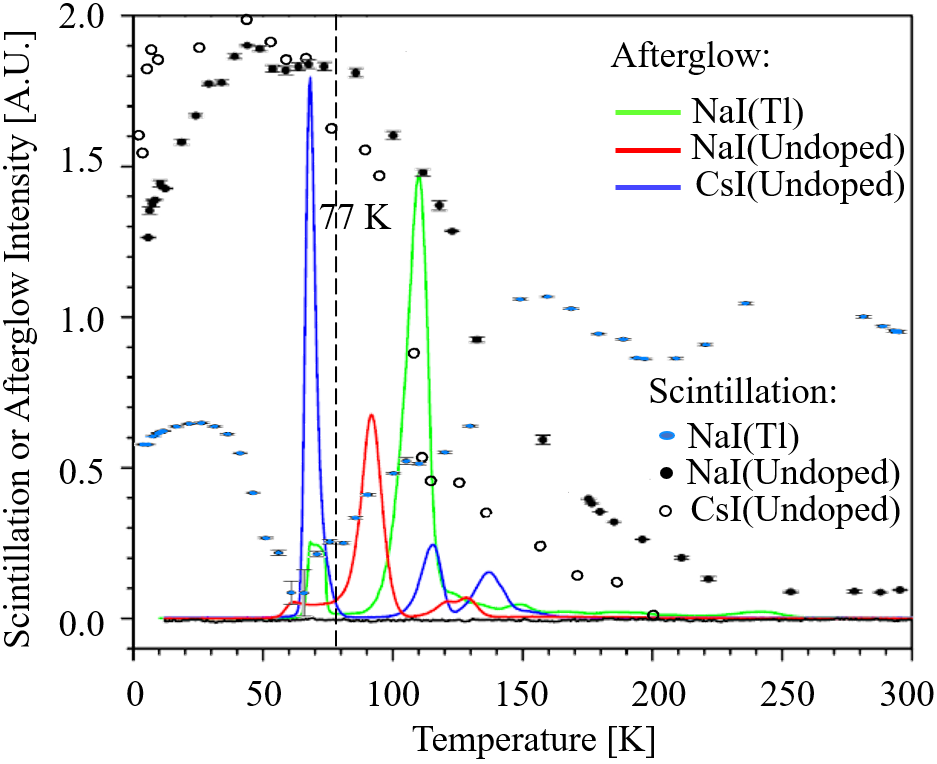}
  \caption{Relative scintillation yields and afterglow rates of various crystals as a function of temperature. The scintillation yield of undoped CsI is taken from Ref.~\cite{Nishimura95}, the yields of undoped NaI and NaI(Tl) are from Ref.~\cite{Sailer12}. The afterglow rates are from Ref.~\cite{derenzo18}.}
  \label{f:ag}
\end{figure}

The energy dispersion among phonons and the two types of excitons dictates the temperature dependence of the light yield of undoped crystals~\cite{Nishimura95}. Their relative scintillation intensities in a wide temperature range are summarized in Fig.~\ref{f:ag}~\cite{Nishimura95, Sailer12, mikhailik15}. For comparison, the relative scintillation intensity of thallium doped NaI~\cite{Sailer12} is plotted in the same figure. A few points are worth emphasizing:

\begin{itemize}
  \item The light yields of undoped crystals peak around 40~K.
  \item Since the curves are relatively flat around their peaks, the yields at the liquid nitrogen temperature are still very close to the maxima.
  \item The yields drop to about 2/3 of the maxima when the temperature goes down to 4~K. The missing energy needs to be harvested from the phonon channel.
  \item The absolute yields of undoped crystals at their maxima are about twice higher than those of doped ones at room temperature.
  \item Cooling existing doped crystals does not result in an increase of the their yields.
\end{itemize}

Compared to direct DM detection experiments deep underground, detectors located at the SNS are much shallower. Afterglows of a crystal induced by energetic cosmic muon events may be a serious concern. As shown in Fig.~\ref{f:ag}, undoped CsI and NaI suffer from afterglow above $\sim$50~K~\cite{derenzo18}.  One can thus maximize the light yield and minimize the afterglow of undoped CsI and NaI by operating them near 40~K. However, such a temperature is less convenient to maintain than 77~K, which can be simply achieved using liquid nitrogen as coolant. Interestingly, there is a valley around 77~K in between the two peaks in the undoped CsI and NaI afterglow distributions. Compromising a bit in the afterglow rate, the prototype can be operated at 77~K for convenience. One can then require a coincident observation of light signals in at least two light sensors to suppress both the afterglow from the crystal and the dark noise from light sensors at the single photon level. A simple toy MC reveals that a 10-ns coincident window between two sensors coupled to the same crystal results in a suppression factor of about $10^{5}$.

\subsection{Scintillation wavelengths and decay times}
Due to completely different scintillation mechanisms, the scintillation wavelengths and decay times of undoped NaI/CsI are quite different from those of NaI/CsI(Tl), as summarized in Table~\ref{t:RT} and Table~\ref{t:LN} for room and liquid nitrogen temperatures, respectively. Undoped NaI is a much faster scintillator than NaI(Tl). It permits a narrower coincidence time window that can further suppress random backgrounds.

\begin{table}[htbp] \centering
  \caption{Scintillation wavelength $\lambda$ and decay time $\tau$ of Tl-doped and undoped NaI, CsI crystals at room temperature.}
  \label{t:RT}
  \begin{tabular}{ccccc}\hline
    Crystal & $\tau$ at $\sim297$~K [ns] & $\lambda$ at $\sim297$~K [nm] \\\hline
    NaI(Tl) & $230\sim250$~\cite{robertson61, eby54, schweitzer83} & $420\sim430$~\cite{Sciver56,Sibczynski12} \\
    CsI(Tl) & 600~\cite{Bonanomi52} & 550~\cite{Kubota88} \\
    undoped NaI & $10\sim15$~\cite{Sciver56,Sciver58,Beghian58} & 375~\cite{West70,Fontana70} \\
    undoped CsI & $6\sim36$~\cite{Kubota88,Schotanus90,Amsler02} & $305\sim310$~\cite{Kubota88,Woody90,Amsler02} \\\hline
  \end{tabular}
\end{table}

\begin{table}[htbp] \centering
  \caption{Scintillation wavelength $\lambda$ and decay time $\tau$ of Tl-doped and undoped NaI, CsI crystals at liquid nitrogen temperature.}
  \label{t:LN}
  \begin{tabular}{ccccc}\hline
    Crystal & $\tau$ at $\sim$77~K [ns] & $\lambda$ at $\sim$77~K [nm]\\\hline
    NaI(Tl) & 736~\cite{Sibczynski12} & $420\sim430$~\cite{Sciver56,          Sibczynski12} \\
    CsI(Tl) & no data & no data \\
    undoped NaI & 30~\cite{Sciver58, Beghian58} & 303~\cite{Sciver56,Sibczynski12}\\
    undoped CsI & 1000~\cite{Nishimura95,Amsler02,csi} & 340~\cite{Nishimura95, Woody90,Amsler02}\\\hline
  \end{tabular}
\end{table}

A thorough measurement of the response of undoped CsI to various radiations in a wide temperature range can be found in Ref.~\cite{clark18}. In general, decay constants change with temperature. However, around 77~K the decay constant of alpha induced scintillation is almost identical to that of photons. A detailed measurement of the decay constant of nuclear recoils in a wide temperature range is needed with the hope of identifying a temperature suitable for particle identification.

\subsection{Energy threshold}
According to Ref.~\cite{csi}, the quantum efficiency of R11065 at 80~K near 300~nm is about 27\%, while the photon detection efficiency of some silicon photomultipliers (SiPM) can already reach 56\% at around 420~nm~\cite{giovanetti17}. By replacing PMTs with SiPM arrays coated with some wavelength shifting material that shifts 313~nm~\cite{Sibczynski12}/340~nm~\cite{Nishimura95, Woody90, Amsler02} scintillation light from undoped NaI/CsI to $\sim$430~nm, it is possible to double the light yield from $25.99 \pm 0.42$~PE/keVee to about 50~PE/keVee. Such a high yield has recently been almost achieved using a combination of a small undoped CsI and a few large-area avalanche photodiodes (LAAPD) after wavelength shifting~\cite{ess19}. Compared to a SiPM, a LAAPD has generally even higher light detection efficiency, but its output signals are too small to be triggered at single-PE level.

\begin{figure}[htbp]
  \includegraphics[width=\linewidth]{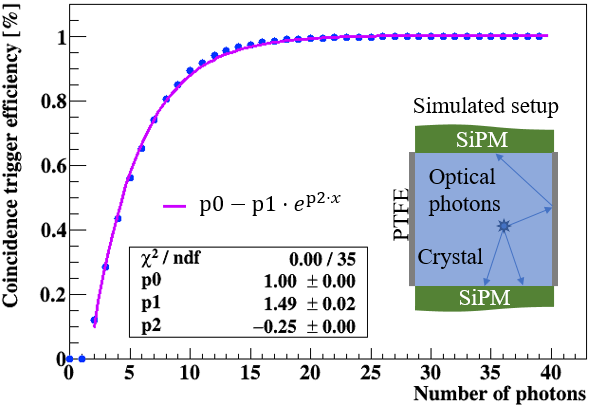}
  \caption{Two-PE coincidence trigger efficiency as a function of the number of emitted photons in the proposed detector.}
  \label{f:thr}
\end{figure}

To estimate the trigger efficiency of a detector module as shown in the inlet of Fig.~\ref{f:thr} that has a light yield of 50~PE/keVee, a toy Monte Carlo simulation was performed as follows:
\begin{itemize}
  \item $n$ photons were generated.
  \item 10\% of them were thrown away randomly, mimicking a 90\% light collection efficiency.
  \item The remaining photons had an even chance to reach individual SiPM arrays, and 56\% of chance to be detected.
  \item If both arrays recorded at least one PE, this simulated event was regarded as being triggered.
\end{itemize}
The value of $n$ changed from 0 to 40. For each value, 10,000 events were simulated. The trigger efficiency was calculated as the number of triggered events divided by 10,000.

Fig.~\ref{f:thr} shows the simulated 2-PE coincidence trigger efficiency as a function of the number of generated photons. An exponential function (purple curve) with three free parameters was fitted to the simulated results (blue dots).  The fitted function was used to convert energy spectra to PE spectra, which is described in detail in the next section.

From Fig.~\ref{f:thr}, one can read a trigger efficiency of \SI{80}{\percent} when there are about 8 photons, which can be converted to about $8\times90\%\times54\%\approx4$~PE, taking into account the light collection efficiency of 90\% and the photon detection efficiency of 56\%. This can be further translated to a threshold of $4/50=80$~eVee, given the 50~PE/keVee light yield.

Assuming a constant quenching factor of 0.08 for NaI and 0.05 for CsI in such a low energy region, the threshold is translated to 1 keV for Na recoils, and 1.6 keV for Cs recoils. Note that the quenching factors assumed here are from measurements with doped crystals~\cite{coherent17, PhysRevC.92.015807}. Given completely different scintillation mechanisms, there is a possibility that scintillation quenching in undoped crystals is less serious than that in doped crystals. For example, a very preliminary investigation~\cite{ess19} suggests a quenching factor of 0.1 for undoped CsI. The assumption here is hence conservative.

\subsection{Source}
The SNS is the world's premier neutron-scattering research facility, producing pulsed neutron beams with intensities an order of magnitude larger than any other currently-operating facility. At its full beam power, about \num{1.5e14} \SI{1}{\GeV} protons bombard a liquid mercury target in \SI{600}{\ns} bursts at a rate of \SI{60}{\hertz}. Neutrons produced in spallation reactions in the mercury target are thermalized in cryogenic moderators surrounding the target and are delivered to neutron-scattering instruments in the SNS experiment hall. The SNS is a user facility and operates approximately two-thirds of the year.

\begin{figure}[htbp]
  \includegraphics[width=\linewidth]{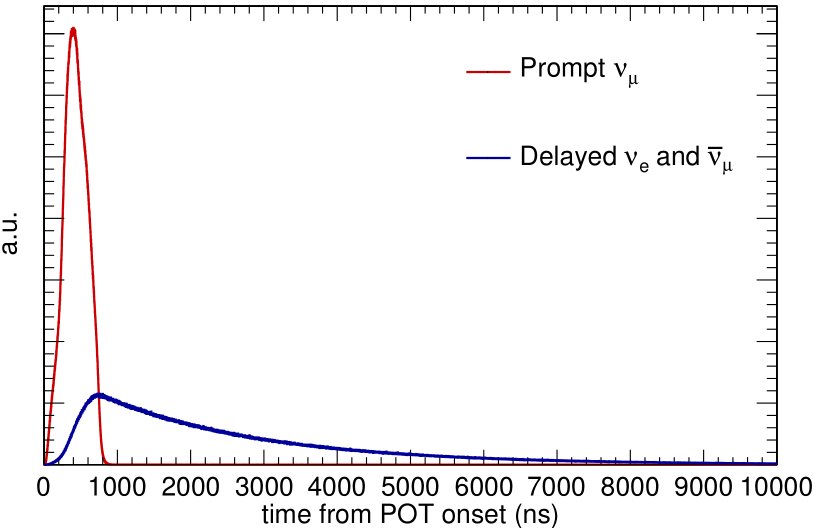}
  \caption{Time structure for prompt and delayed neutrinos due to the 60-Hz pulses~\cite{coherent18}. ``Prompt'' refers to neutrinos from pion decay, and ``delayed'' refers to neutrinos from muon decay.} \label{f:vd}
\end{figure}

As a byproduct, the SNS also provides the world's most intense pulsed source of neutrinos in an energy region of specific interest for particle and nuclear astrophysics. Interactions of the proton beam in the mercury target produce $\pi^+$ and $\pi^-$ in addition to neutrons. These pions quickly stop inside the dense mercury target. Most of $\pi^-$ are absorbed. In contrast, the subsequent $\pi^+$ decay-at-rest (DAR) produces neutrinos of three flavors. The produced neutrino energy spectra are shown in the left plot in Fig.~\ref{f:vd}.

\begin{figure}[htbp]\centering
  \includegraphics[width=\linewidth]{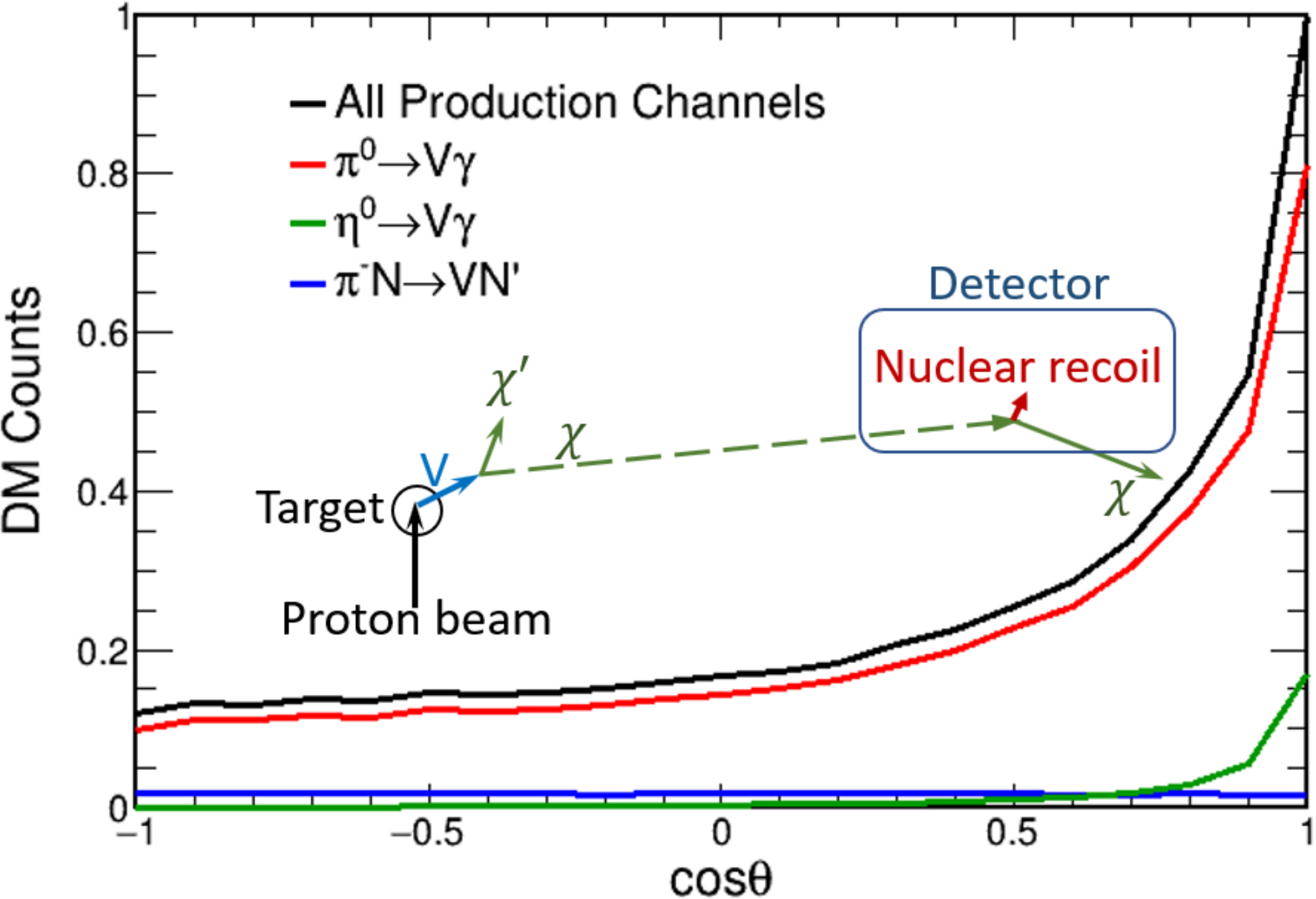}
  \caption{Angular distributions of DM particles created from various channels. The inlet shows  the production and detection mechanism.}
  \label{f:pro}
\end{figure}

DM particles are another possible byproduct. In the forward direction, the dominant DM production channel is the decay of $\pi^0$/$\eta^0$ particles on the fly, while the nuclear absorption of $\pi^-$ particles may produce portal particles isotropically. The portal particle would subsequently decay to a pair of light DM particles, $\chi^\dagger\chi$, either of which may interact with a detector as shown in Fig.~\ref{f:pro}.

The sharp SNS beam timing structure is highly beneficial for background rejection and precise characterization of those backgrounds not associated with the beam~\cite{Bolozdynya:2012xv}, such as those from radioactive impurities in a crystal.  Looking for beam-related signals only in the \SI{10}{\micro\s} window after a beam spill imposes a factor-of-2000 reduction in the steady-state background. Note that DM signals only appear in the prompt neutrino window. If nonstandard neutrino interactions have negligible contribution, the CEvNS measurement in the delayed neutrino window can be used to constrain the CEvNS background in the prompt window for DM search~\cite{dutta20}. A simultaneous fit of prompt and delayed events utilizing both the time and spectral information will mitigate the systematic uncertainty on the prompt CEvNS background~\cite{lardm19, dutta20}.

\subsection{Location}
\begin{figure}[htbp]\centering
  \includegraphics[width=\linewidth]{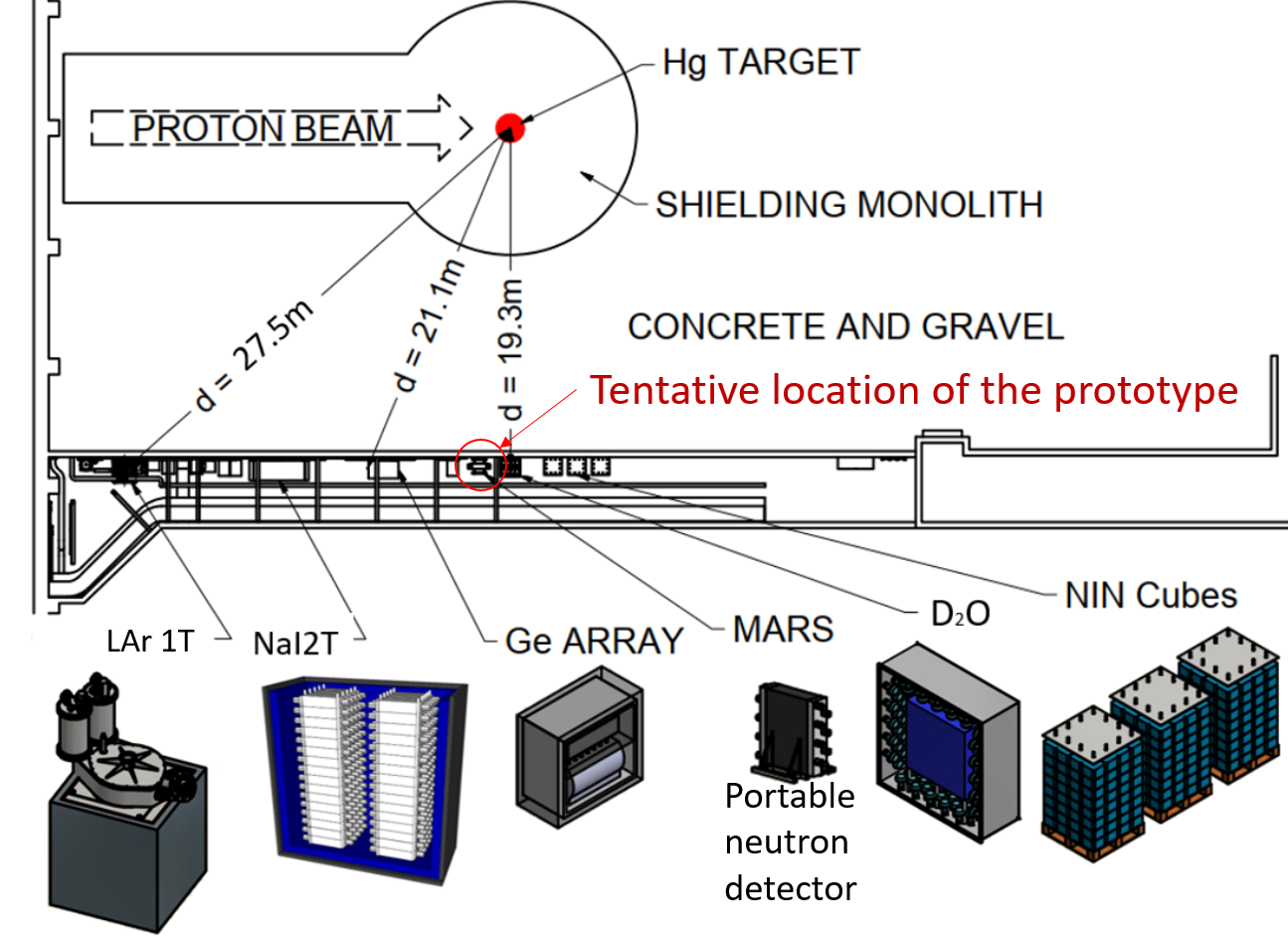}
  \caption{Possible location of the proposed detector in the Neutrino Alley at the SNS.}
  \label{f:alley}
\end{figure}

The COHERENT Collaboration occupies the ``Neutrino Alley'' located $\sim$20~m from the mercury target with contiguous intervening shielding materials and overburden eliminating almost all free-streaming pathways for fast neutrons which dominate beam-related backgrounds.  The presumed prototype can be located very close to the previous 14~kg CsI(Na) detector as shown in Fig.~\ref{f:alley}.

\subsection{Sensitivity to low-mass dark matter}
Given the operational parameters outlined in the previous sections, the sensitivity of the prototype detector placed 19.3 meters away from the SNS target for low-mass DM detection was estimated.

Two classes of dark matter portal particles can be constrained by such an experiment: a vector portal particle kinetic mixing with a photon, and a leptophobic portal particle coupling to any SM baryon.  In addition to the portal and the DM particle masses, $m_V$ and $m_\chi$, the vector portal model has two coupling constants as free parameters, $\epsilon$ and $\alpha'$,  while the leptophobic parameter depends on a single $\alpha_B$.  The parameters of the vector portal model can be conveniently compared to the cosmological relic density of dark matter through the dimensionless quantity, $Y=\epsilon^2\alpha'(m_\chi/m_V)^4$~\cite{Izaguirre:2015yja}, which can easily be compared to results from direct detection experiments.  The sensitivity to the leptophobic portal of the assumed detector is of great interest compared to beam dump experiments, which are frequently most sensitive to $\nu$-$e$ elastic scattering~\cite{PhysRevD.63.112001,deNiverville:2011it}, and are incapable of testing this model.

\begin{figure}[htbp] \centering
  \includegraphics[width=\linewidth]{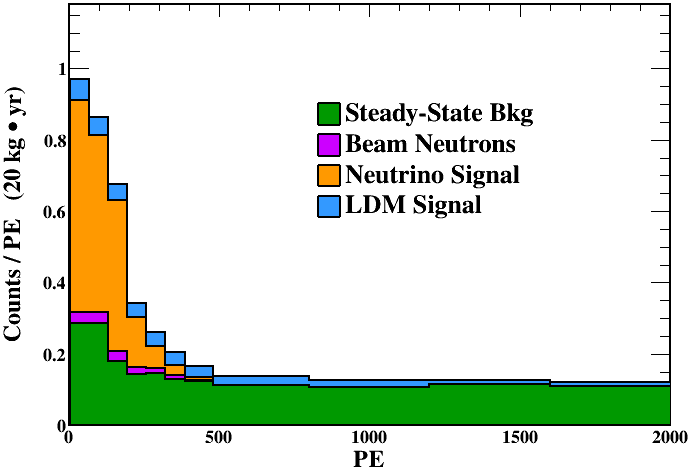}
  \caption{Energy spectra of the proposed detector at the SNS in the prompted neutrino window ($0 \sim 6 \mu$s) with an exposure of 20 kg$\cdot$year.}
  \label{f:ldm}
\end{figure}

The BdNMC event generator~\cite{den17} was used to determine the energy spectra of Na and I recoils in the assumed detector, parameterized by the dark matter and portal particle masses~\cite{lardm19}. Assuming a constant nuclear recoil quenching factor of 0.08, the generated Na and I recoil energy spectra were converted to visible energy spectra in keVee. The 50 PE/keVee system light yield was translated to the crystal's intrinsic light yield of $50/56\%/90\%\approx100$~photons/keVee, which was used to convert visible energy spectra to number-of-photon spectra. A simple Poisson smearing of the number of photons was applied to the latter. At last, the trigger efficiency function fitted to Fig.~\ref{f:thr} was applied to convert the number-of-photon spectra to PE spectra, which were summed and shown as the blue histogram stacked on top of others in Fig.~\ref{f:ldm} labeled as ``LDM Signal''. The total number of LDM events integrated over the whole spectrum at $Y=2.6\times 10^{-11}$ and $m_\chi=10$~MeV is about 44.

The largest component in Fig.~\ref{f:ldm} colored in orange and labeled as ``Neutrino Signal'' are the calculated CEvNS spectrum with the detector responses folded in. The total number of events is about 218 in the $0 \sim 0.8 \mu$s prompt neutrino window. Additional 663 CEvNS events can be detected in the delay window ($0.8 \sim 6~\mu$s), which were used to constrain the uncertainty of the orange spectrum in Fig.~\ref{f:ldm}.  The bottom two histograms labeled ``Beam Neutrons'' and ``Steady-State bkg'' are the SNS beam related and unrelated background spectra measured by the COHERENT CsI(Na) detector ~\cite{coherent17}. Since the proposed detector has a much lower threshold, there is no measurement of the two backgrounds below 40~PE. The rates of the two were assumed to be flat below 40~PE.

\begin{figure}[htbp] \centering
  \includegraphics[width=\linewidth]{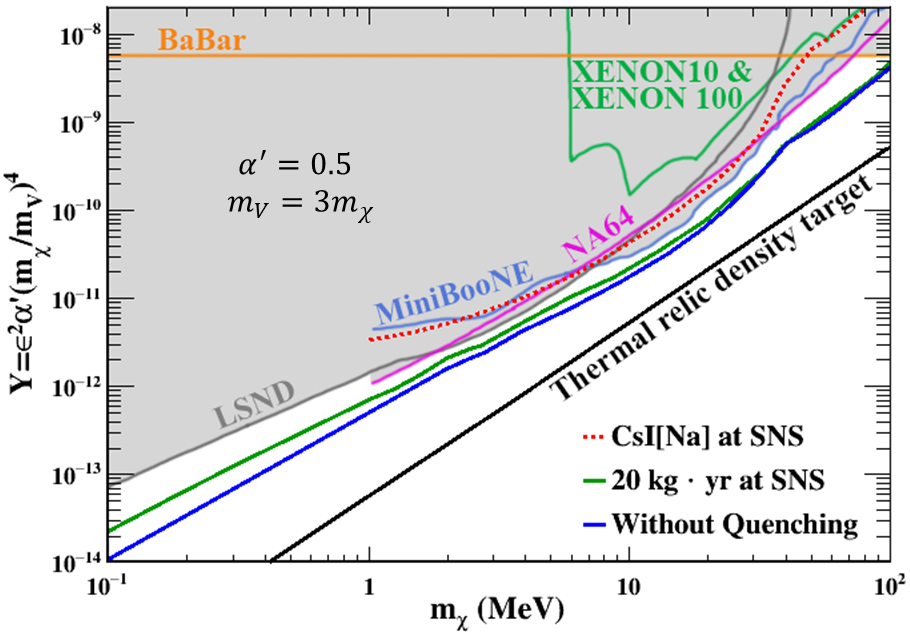}
  \caption{Predicated 90\% sensitivity to low-mass dark matter production parameters in case of the vector portal theory assuming $\alpha' = 0.5$ and $m_V = 3 m_\chi$.}
  \label{f:y}
\end{figure}

\begin{figure}[htbp] \centering
  \includegraphics[width=\linewidth]{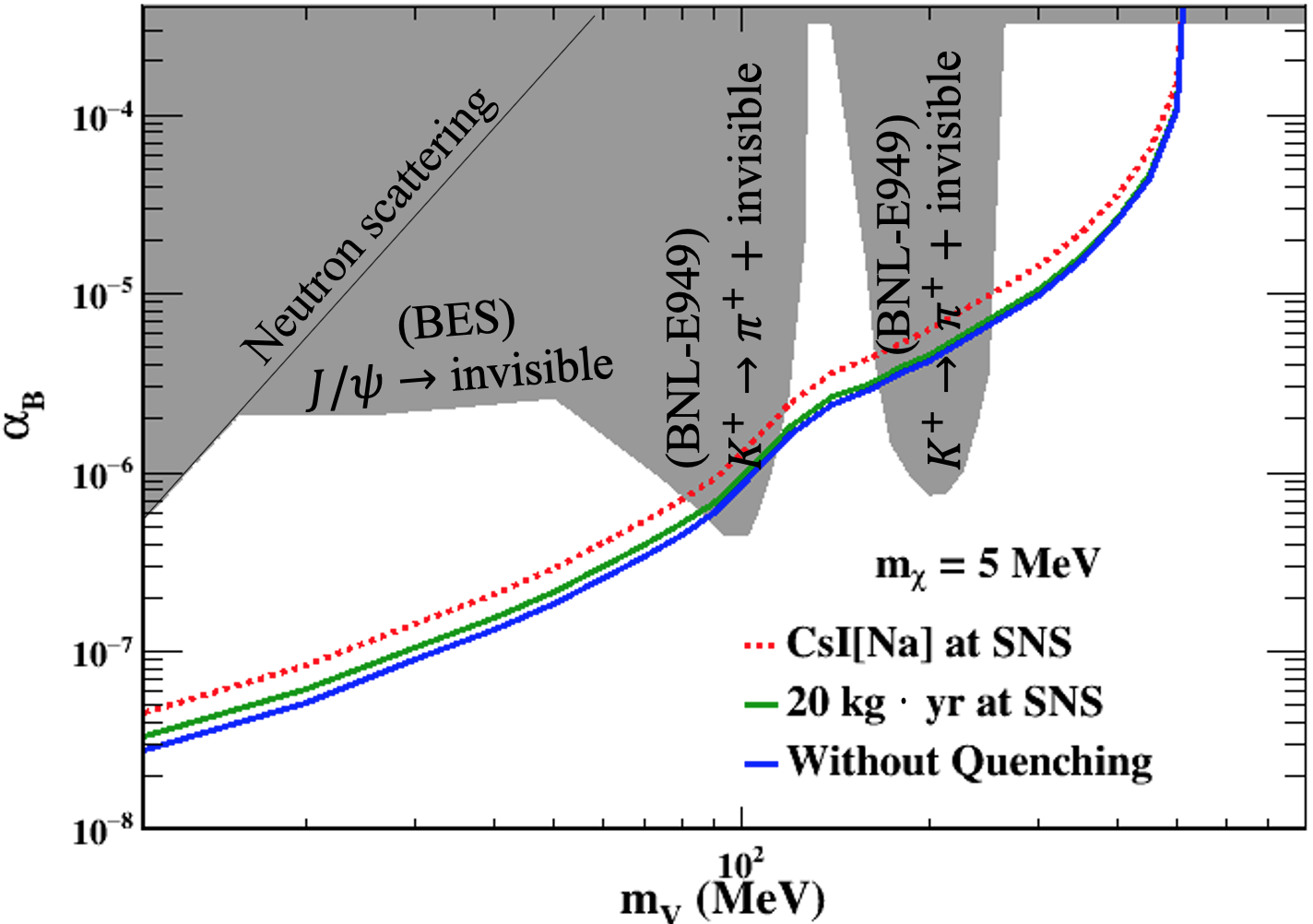}
  \caption{Predicated 90\% sensitivity to low-amss dark matter production parameters in case of the baryonic portal theory.}
  \label{f:alpha}
\end{figure}

For each $m_\chi$ and $m_V$, the minimum dark matter coupling constants that are inconsistent with the Asimov prediction~\cite{cowan11} was calculated taking into account systematic uncertainties as described in detail in Ref.~\cite{lardm19}. The results are shown in Fig.~\ref{f:y} and \ref{f:alpha} for an exposure of a 10~kg crystal for 2 years of data taking. For comparison, exclusion curves from other experiments~\cite{deNiverville:2011it, PhysRevD.63.112001, Auerbach:2001wg, Aubert:2008as, PhysRevD.98.112004, Gninenko:2019qiv, na6419} are overlaid with different colors and labels. The ``Thermal relic density target'' line indicates the model parameters, where DM interactions with visible matter in the early hot Universe, explain the DM abundance today. Since parameters below it would over-produce DM in freeze-out, it is the target line to verify the model.

The nuclear quenching factor of undoped NaI has not been measured. Small or no quenching were observed in undoped CsI for $\alpha$ radiation compared to $\gamma$ radiation~\cite{Hahn53, Schotanus90}. A very preliminary measurement of the nuclear quenching factor of an undoped CsI gives a value of 0.1~\cite{ess19}. Detailed measurement of the nuclear quenching factors for both undoped NaI and CsI is planned. For the purpose of sensitivity estimation, two extreme cases are considered. The red curves in Fig.~\ref{f:y} and \ref{f:alpha} correspond to a constant quenching factor of 0.08. The blue ones are with no quenching at all. The real sensitivity curve should lay in between.

\section{Conclusion}
The sensitivity of a 10~kg prototype detector based cryogenic inorganic scintillating crystals to detect low-mass dark matter particles produced at the Spallation Neutron Source at the Oak Ridge National Laboratory was investigated. After two years of data taking, the presumed detector can be used to explore large areas of phase space, which have not been covered by past underground and accelerator-based experiments.  The presumed detector consists of a 10~kg undoped CsI or NaI crystal directly coupled to SiPM arrays operated at liquid nitrogen temperature, the key technical advantage of which is the much higher light yields of undoped crystals at 77~K compared to those of doped ones at room temperature. As the first step to verify the feasibility of the proposed technique, the light yield of an undoped CsI crystal directly coupled to two PMTs at about 77~K was measured. The cylindrical crystal has a diameter of 3~inches, a height of 5~cm, and a mass of 1.028~kg. A light yield of $26.0\pm0.4$~PE/keVee was achieved. This was $\sim30\%$ higher than a previous measurement using a much smaller crystal~\cite{csi}, $\sim160\%$ higher than the highest light yield achieved by DAMA~\cite{dama18}, $\sim68\%$ higher than the highest light yield achieved by COSINE~\cite{cosine19}. To the authors' best knowledge, this is the highest in the world achieved with a crystal more than 1~kg.

\begin{acknowledgements}
  This work is supported by the NSF award PHY-1506036, and the Office of Research at the University of South Dakota. Computations supporting this project were performed on High Performance Computing systems at the University of South Dakota, funded by NSF award OAC-1626516.
\end{acknowledgements}

\bibliography{ref}
\bibliographystyle{spphys}

\end{document}